\documentclass[%
 reprint,
 amsmath,amssymb,
prb,
]{revtex4-2}

\usepackage{graphicx}
\usepackage{dcolumn}
\usepackage{bm}
\usepackage{hyperref}
\usepackage{subfig}

\begin{document}

\preprint{APS/123-QED}

\title{Linear and nonlinear properties of a compact  high-kinetic-inductance WSi multimode resonator}

\author{Naftali Kirsh}
\affiliation{Racah Institute of Physics, the Hebrew University of Jerusalem, Jerusalem, 91904 Israel}
\author{Elisha Svetitsky}
\affiliation{Racah Institute of Physics, the Hebrew University of Jerusalem, Jerusalem, 91904 Israel}
\author{Samuel Goldstein}
\affiliation{Racah Institute of Physics, the Hebrew University of Jerusalem, Jerusalem, 91904 Israel}
\author{Guy Pardo}
\affiliation{Racah Institute of Physics, the Hebrew University of Jerusalem, Jerusalem, 91904 Israel}
\author{Ori Hachmo}
\affiliation{Racah Institute of Physics, the Hebrew University of Jerusalem, Jerusalem, 91904 Israel}
\author{Nadav Katz}
\affiliation{Racah Institute of Physics, the Hebrew University of Jerusalem, Jerusalem, 91904 Israel}

\date{\today}

\begin{abstract}
The kinetic inductance (KI) of superconducting devices can be exploited for reducing the footprint of linear elements as well as for introducing nonlinearity to the circuit. We characterize the linear and nonlinear properties of a multimode resonator fabricated from amorphous tungsten silicide (WSi) with a fundamental frequency of \(f_1 = 172\) MHz. We show how the multimode structure of the device can be used to extract the different quality factors and to aid the nonlinear characterization. In the linear regime the footprint is reduced by a factor of \(\sim 2.9\) with standard lateral dimensions with no significant degradation of the internal quality factor compared to a similar Al device . In the nonlinear regime we observe self positive frequency shifts at low powers which can be attributed to saturation of tunneling two-level systems. The cross mode nonlinearities are described well by a Kerr model with a self-Kerr coefficient in the order of \(|K_{11}|/2\pi \approx 1.5\times10^{-7}\) Hz/photon. These properties together with a reproducible fabrication process make WSi a promising candidate for creating linear and nonlinear circuit QED elements.  
\end{abstract}

\maketitle
\section{Introduction}
The increasing complexity of superconducting circuits\cite{arute2019quantum,jurcevic2021demonstration} requires miniaturization of the circuit elements. For resonators used for qubit readout \cite{Blais2004}, coupling \cite{majer2007coupling,jurcevic2021demonstration} and as Purcell filters \cite{reed2010fast} this can be done by increasing the inductance.  A promising direction is the usage of high kinetic inductance (KI) materials. Kinetic inductance is at the heart of various types of kinetic inductance detectors (KIDs)  \cite{DayMKID,mazin2020superconducting,TKID}. In addition, linear high kinetic inductance can also be used to produce high impedance superinductors\cite{superIndGrAl}. Furthermore, the kinetic inductance of superconducting thin films depends nonlinearly on the circulating current according to \cite{Zmuidzinas2012}
\begin{equation} \label{eq:KI}
	L_k = L_{k,lin.}\left(1+\left(\frac{I}{I_*}\right)^2\right), 
\end{equation}where \(I_*\) is of the order of the critical current\cite{semenov2020effect}.
This nonlinearity can be exploited for applications such as traveling wave amplifiers \cite{vissersKI3WM,malnouKI3WM_new, goldstein2020four} and even qubits\cite{grAlQubit}.   Several high KI materials were employed, such as granular Al\cite{grAlLoss,rotzinger2016aluminium} , disordered Al \cite{disorderedAl}, NbN\cite{niepce2019high,xing2020compact}  and TiN\cite{TiN}. 
\newline

Here we use amorphous tungsten silcide (WSi) as a high KI material. WSi is an amorphous superconductor in which \(T_c\) and the normal state resistance can be controlled by the amount of silicon doping and thickness\cite{kondo1992superconducting,WSi_MKID,WSi_char_SPD,SNPD}. WSi was used for single photon detection due to its structural homogeneity and large hotspot size\cite{SNPD,SNPD_93}.  The usage of WSi for microwave\cite{WSi_MKID} and thermal KIDs\cite{TKID} in the x-ray regime was also investigated.  
Here we demonstrate a compact multimode resonator made of a WSi film.  In addition to the reduction of footprint due to the linear kinetic inductance, the nonlinearity of the device results in an intermode  coupling which we use as an aid for the device characterization. 
Our device has a low loss and as shown below can be used in both the linear and nonlinear regimes. The fabrication method is reproducible and was also used to create a traveling-wave amplifier\cite{goldstein2020four},  compact microwave photonics networks \cite{goldstein2021compact} and standard notch-type resonators\cite{suppMaterial}.  In order to isolate the effects of the kinetic inductance from the geometrical inductance we also fabricated an Al device with identical geometrical dimensions (except for the metal's thickness). Since the capacitance and magnetic (geometrical) inductance depend almost only on geometry and the substrate dielectric constant \cite{Goppl2008} they should be identical for both devices.    

\section{Fabrication}
Thin films of WSi are deposited via DC magnetron sputtering. WSi sputter targets are commercially available, and we use a W\textsubscript{0.55}Si\textsubscript{0.45} target with a 2 inch diameter supplied by Kurt J. Lesker. WSi is a poor conductor, and utilizing sputter settings similar to those for Al and Nb processes heats the target to the point of melting the indium bonds on its backside. We find that target heating is mitigated by using a low sputtering power of 42 W and a resulting deposition rate of 5 nm/min.

\begin{figure}
	\includegraphics[width=\linewidth,trim={0cm 1.05cm 0cm 0cm},clip]{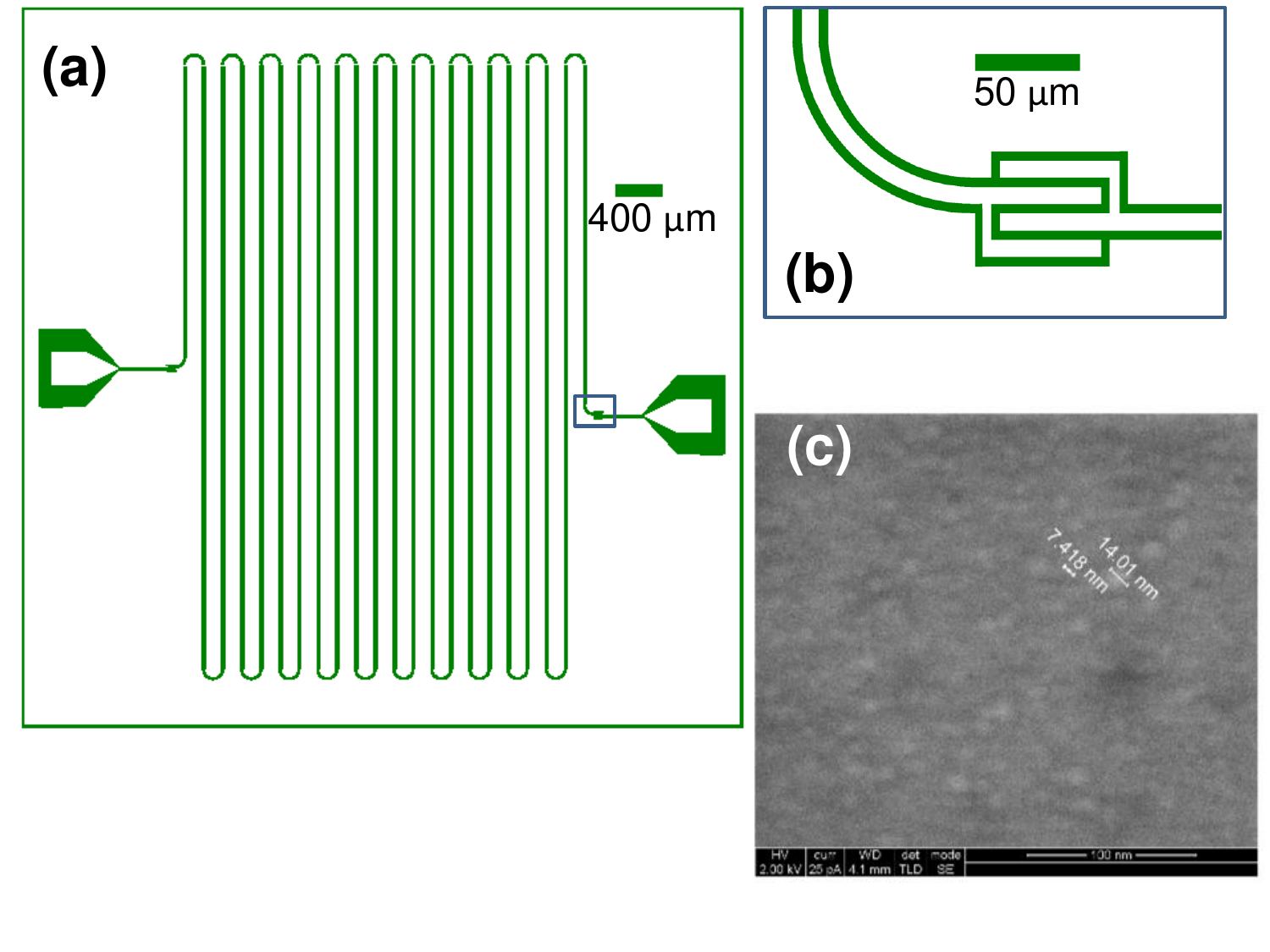} 
	\caption{Multimode resonator device. (a)-(b) Sketch of the device, (b) shows the coupling to the feedline via an interdigitated capacitor. White: metal trace and ground. Green: exposed Si substrate. (c) SEM image of the surface of a sputtered 30 nm WSi film. Grains of order 10 nm are visible.}
	\label{fig:fab}
\end{figure}

We deposit 30 nm of WSi on a high resistivity ($\rho > 10^4  \;  \Omega \cdot cm$) Si substrate. A meandered coplanar waveguide resonator of length 11.6 cm, central strip width 8 $\mu$m, and a spacing to ground of 5 $\mu$m  is defined via optical lithography. The resonator is capacitively coupled to a feedline at either end with an interdigitated capacitance with 4 fingers  50 \(\mu m\)  long each.  See Fig. \ref{fig:fab}. The exposed WSi is removed with a tungsten etchant supplied by Sigma Aldrich, with an etch rate of \(\sim 2-3\) nm/sec at room temperature. The etchant does not attack PMMA, and is therefore compatible also with electron beam lithography processes.
An identical resonator is fabricated with a film of 80 nm Al.
Fig. \ref{fig:fab}(c) shows SEM images of the surface of a 30 nm WSi film, revealing a weakly observable graininess on the order of 10 nm.  The insignificance of the grains enables patterning of \(\sim 0.25-0.5\,\mu m\)  nanowires from sputtered WSi films in which the KI agrees well with the geometric scaling\cite{goldstein2021compact}.

The devices are mounted in their respective holders and bolted to the base plate of a dilution refrigerator, where they are measured at a base temperature of 20 mK.  Linear characterization is done using a vector network analyzer. For the nonlinear mode coupling measurements we combine the network analyzer probe signal  with a CW tone from a microwave signal generator.

\section{Linear characterization}
\subsection{Kinetic inductance fraction and footprint reduction}
\begin{figure}
    \includegraphics[width=\linewidth,trim={0cm  7.5cm 2.5cm 0cm},clip]{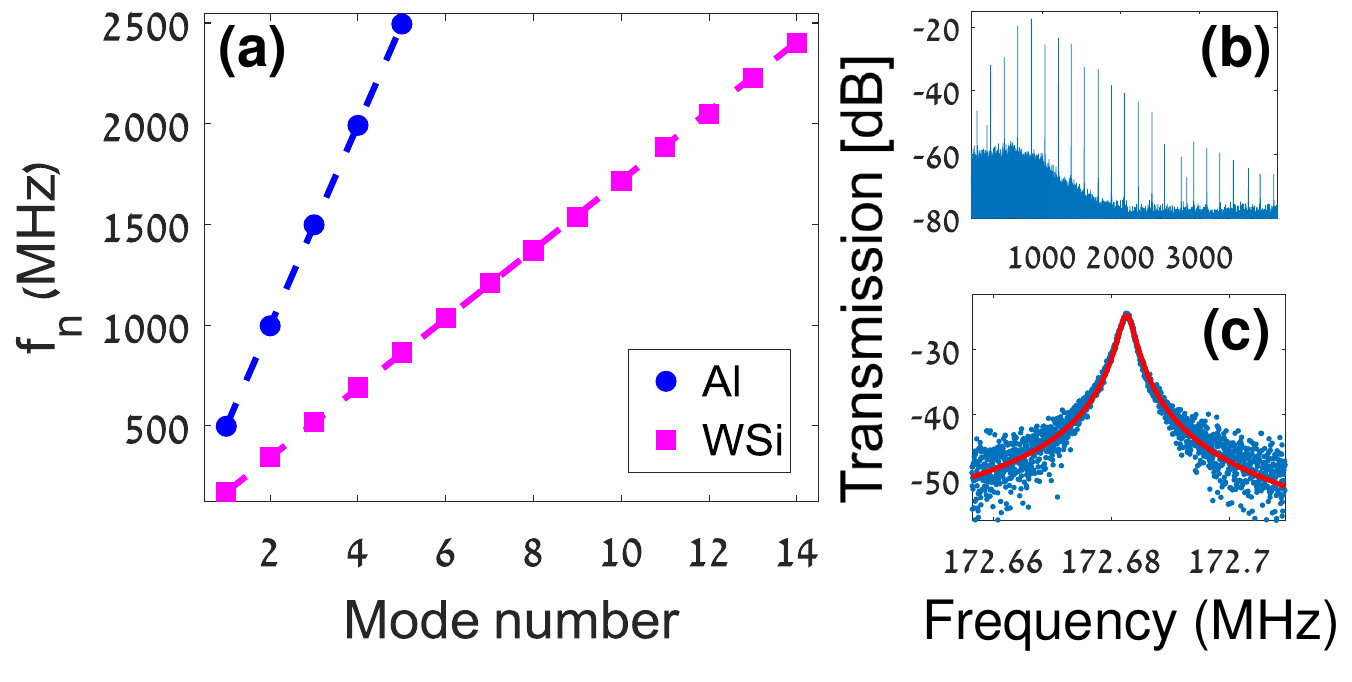}
    \caption{Reducing the footprint of the multimode resonator. (a) Multimode spectra of two geometrically identical Al and WSi multimode resonators. The lines are linear fits for the modes' frequencies. The WSi device obtains a frequency spacing smaller by  a factor of \(\approx 2.9\)  from the Al device using the same length. (b) Raw transmission spectrum of the WSi device, showing the first 23 modes. The profile is affected by the amplifiers' frequency response. The lower peaks at 300 and 2844 MHz are probably an electronic feed-through and a slot mode. (c) Magnitude of a typical transmission data around resonance. The red line is the magnitude of the fit to the complex data\cite{mazin2005microwave,Probst2014}.} \label{fig:spectra}
\end{figure}

Our devices are two geometrically identical 11.6 cm long meandered coplanar waveguides terminated at both ends by interdigitated capacitors, one made of standard Al and the other from WSi as described above.  As can be seen in Fig. \ref{fig:spectra}(b) the \(\lambda/2\)  boundary conditions result in a transmission spectrum of  equally spaced peaks at integer multiples of the fundamental frequency \(f_1 \equiv \tilde{c}/2l\), where \(\tilde{c} = \left(L_l C_l \right)^{-1/2}\) is the phase velocity of wave propagation along the waveguide , with \(L_l\) \(\left(C_l\right)\) the inductance (capacitance) per unit length and \(l\) is the length of the resonator \cite{Goppl2008}.  
Fig.\ref{fig:spectra}(a) shows linear fits with the function \(f_n^{Al,WSi} = f_1^{Al,WSi} \cdot n\)  from which we extract the fundamental frequencies. Using the frequencies of both devices and neglecting the kinetic inductance of the Al\cite{gao2006experimental} we can obtain the kinetic inductance fraction of the WSi device
\begin{equation} \label{eq:alpha}
	\alpha \equiv \frac{L_k}{L_k+L_g} = 1-\frac{L_g}{L_k+L_g} = 1-\left(\frac{f_1^{WSi}}{f_1^{Al}}\right)^2,
\end{equation}
where the fact that the geometrically-dependent capacitance is identical for both devices was used.  The fits yield \(f_1^{Al} \approx 499 \) MHz and \(f_1^{WSi} \approx 172 \)  MHz from which we obtain \(\alpha = 0.88\left(0.02\right)\) or equivalently \(L_k \approx 7.3 L_g\).  Since the required length for a given fundamental frequency is \(l=\tilde{c}/2f_1\) the reduction of length in our case is by a factor of \(\sqrt{1-\alpha} \approx 0.35\), i.e. the footprint is smaller by \(\sim 2.9\). Transport measurements of our WSi films give \(R \approx 98\,\Omega/\square\) and \(T_c \approx 4.75\) K \cite{suppMaterial} which, assuming a BCS superconductor, yields \cite{tinkham2004introduction} \(L_k \approx 0.03\) nH\(/\square\) in agreement with the extracted \(\alpha\) \cite{suppMaterial}. 
\subsection{Quality factors}
\begin{figure}
    \includegraphics[width=\linewidth,trim={0cm  8.15cm 0cm 0cm},clip]{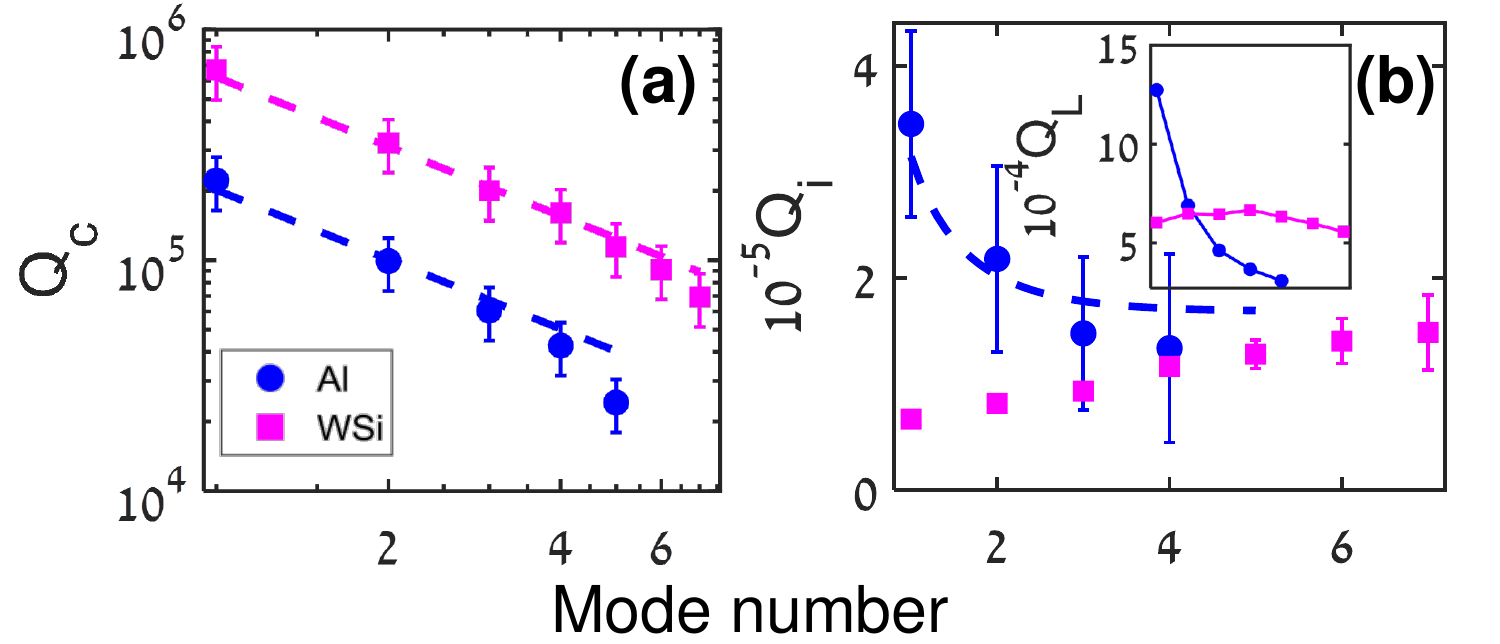}
    \caption{Quality factors of the devices for different modes. (a) Coupling quality factors for the Al (blue) and WSi (magenta) devices (log-log plot). The dashed lines are fits for the expected 1/n dependence Eqn. \ref{eq:Q_c}. (b) Extracted internal factors for the Al (blue) and WSi (magenta) devices. \(Q_i\) is lower for the WSi device probably because a larger fraction of the field energy is stored in the surfaces’ oxides. The dashed line is a fit to a model of thermal saturation of loss-inducing tunneling two-level systems Eqn. \ref{eq:TLS_temp}. The lower internal Q at low modes for the WSi device is unexpected and might be a result of coupling to unwanted slot modes (see Fig.\ref{fig:spectra}(b)). The inset in panel (b) shows the loaded (total) quality factor which is dominated by internal loss (coupling) for the WSi (Al) device. Lines are guides to the eye. 
    } \label{fig:Qs}
\end{figure}

We next turn to characterize the internal losses of the devices. The loaded quality factors \(Q_L\) of the different modes are fitted from the complex transmission data using standard methods \cite{mazin2005microwave, Probst2014}. Since for our geometry the off resonance transmission should ideally vanish we cannot use it for normalization as done with notch-type resonators \cite{Probst2014}. Instead we use the fact that the peak transmission in our case satisfies \(S_{21}^{max} = \frac{Q_L}{Q_c}\), where \(Q_c\) is the coupling quality factor \cite{Goppl2008}.  We note that the coupling quality factor should depend on the mode number \(n\) according to \cite{mazin2005microwave}
\begin{equation} \label{eq:Q_c}
	Q_c = \frac{n\pi}{4 Z_0 Z_L\left(\omega_n C_c\right)^2} = \frac{\pi}{n \cdot 4 Z_0 Z_L\left(\omega_1 C_c\right)^2} \equiv \frac{Q_{c,1}}{n},
\end{equation}
where \(C_c\) is the mode-independent coupling capacitance, \(Z_L=\sqrt{\frac{L_l}{C_l}}\) is the resonator characteristic impedance and \(Z_0\) is the line characteristic impedance approximated to \(50\,\Omega\) neglecting the short WSi feedline. 
This requires careful calibration of the data, since the transmission of the measurement line is frequency dependent (see Fig. \ref{fig:spectra}(b)). We perform this calibration by shorting the input and output lines using a standard SMA cable.  In order to compensate for uncertainties in this method we do not use the raw value \(Q_c = S_{21}^{max} \cdot {Q_L}\) but instead fit to Eqn. \ref{eq:Q_c}, resulting in \(Q_{c,1}^{Al} = 2(0.36)\times10^5\)  for Al, which agrees with the geometry of the design, and \(Q_{c,1}^{WSi} = 6.2(0.54)\times10^5\)  for WSi. See Fig. \ref{fig:Qs}(a).  From Eqns. \ref{eq:Q_c} and \ref{eq:alpha}  we can get \(\frac{Q_{c,1}^{WSi}}{Q_{c,1}^{Al}} = \sqrt{\frac{L_g}{L_g+L_k}}\left(\frac{f_1^{Al}}{f_1^{WSi}}\right)^2 = \frac{1}{\sqrt{1-\alpha}} = 2.9(0.2)\), which agrees with the ratio of the extracted values. Using the values of \(Q_c\) we can extract the internal quality factors as \(Q_i^{-1} = Q_L^{-1}-Q_c^{-1}\). The results are shown in Fig. \ref{fig:Qs}(b) for a probe power of \(\sim -83\) dBm. The dominant loss mechanism for superconducting resonators at low temperature are tunneling two-level systems (TLSs) which can dissipate energy from the electric field to phonons or quasi particles \cite{phillips1987two,martinis2005decoherence,muller2019towards}. In this model the expected temperature dependence of the internal quality factor due to TLSs saturation is \cite{gaoThesis,Pappas2011}
\begin{equation} \label{eq:TLS_temp}
	\frac{1}{Q_i} = \frac{1}{Q_i^0}\tanh\left(\frac{\hbar\omega_n}{2k_BT}\right),
\end{equation}
where \(Q_i^0\) is the low temperature, weak field internal quality factor.  As can be seen in Fig. \ref{fig:Qs}(b) the Al device behavior is well fitted by the theory, while for the WSi device it seems that \(Q_i\)  hardly changes with frequency or even tends to increase at high frequencies. This might imply that other temperature-independent loss mechanisms exist for this device, but might also be a result of coupling to a slot mode at low frequencies (see Fig. \ref{fig:spectra}(b)).  Comparing modes which are close in frequency we have for the 2nd mode of the Al device \(Q_{i,2}^{Al} = 2.2(0.9)\times10^5\) while for the sixth mode of the WSi device \(Q_{i,6}^{WSi} = 1.4(0.2)\times10^5\). This minor degradation of the quality factor is expected when the geometrical dimensions are smaller since a larger fraction of the field energy is stored in the surfaces’ oxides which contain the tunneling TLSs \cite{Gao2008}. 
The noise properties of a traveling-wave amplifier fabricated from WSi (with an additional aSi layer) were measured to be close to the quantum limit \cite{goldstein2020four}, indicating that TLSs at WSi's oxide layers don't contribute unusually to the noise characteristics of the device.   
We notice that since the stored energy depends on the integral of the electric field \(\textbf{e}\) over the volume \(V\) as \(E_{stored} = N\hbar\omega_i \propto \int_V|\textbf{e}|^2 dV\) the relatively low resonance frequencies and long device result in electric field strengths which might not fully saturate the TLSs even at high photon numbers. 
Additional surface treatment may further reduce the loss.   

\section{Nonlinear characterization}
\subsection{Self nonlinearity}
\begin{figure}
    \includegraphics[width=\linewidth,,trim={0cm  0cm 4cm 0cm}]{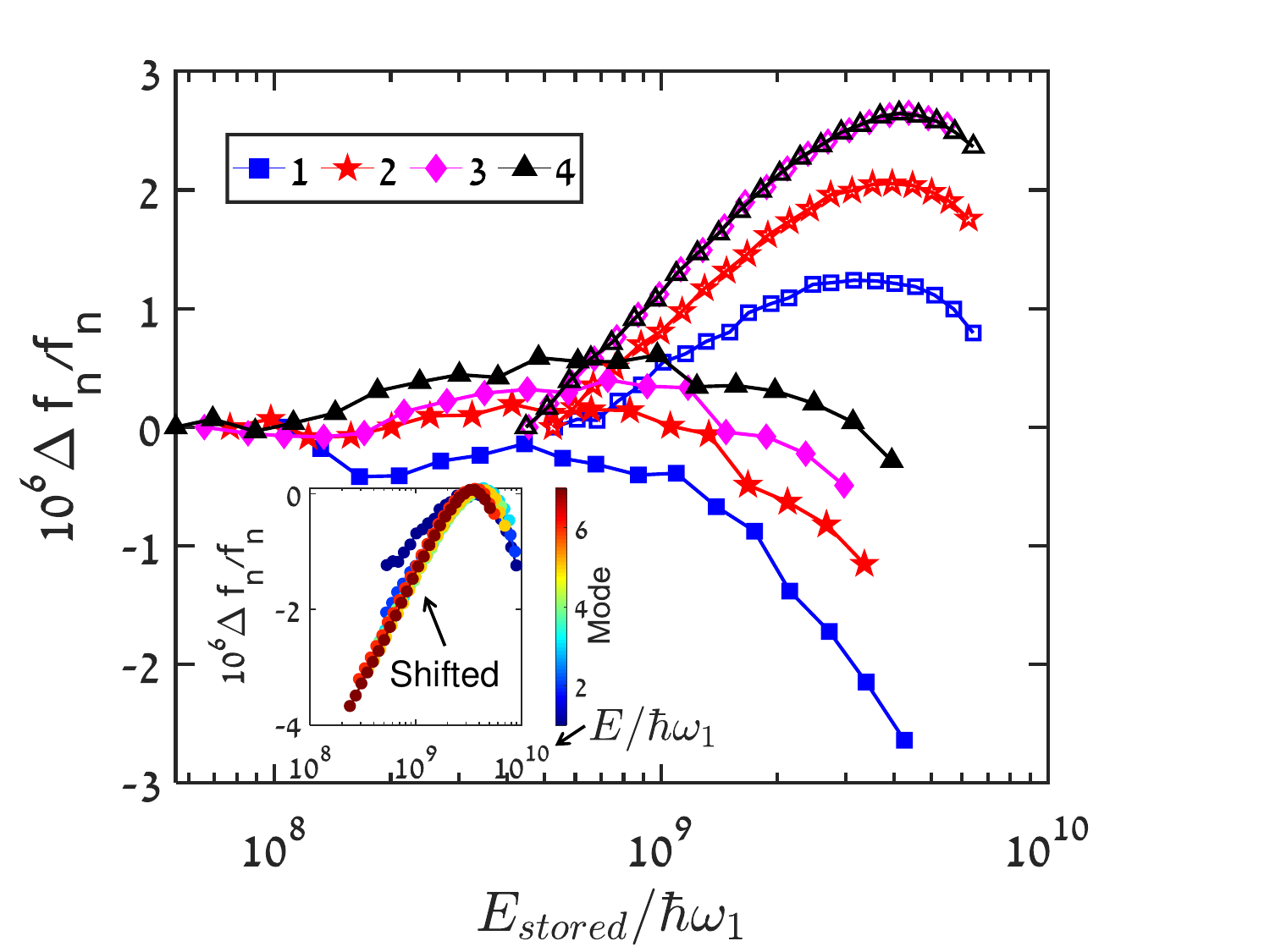}%
    \caption{Self nonlinearity frequency shifts.  Differential frequency shift for the first 4  modes (denoted by different marker shapes as detailed in the legend) of the WSi device as a function of the stored energy (in units of 1st mode photons) for two different cool-downs.  Empty (full) markers are measurements from the first (second) cool-down. Lines are guides to the eye.  The inset shows the (vertically shifted) frequency shifts of the first 7 modes measured at the first cool-down. The positive frequency shifts hinder the evaluation of the bare self-Kerr coefficients.  
    } \label{fig:selfKerr}
\end{figure}

As described in Eqn. \ref{eq:KI}, at high powers the kinetic inductance has a nonlinear dependence on current. This dependence results in a shift of the resonance frequency which is linear in the stored energy, i.e. a Kerr nonlinearity \cite{yurke2006performance}. We can approximate the shift of mode \(n\) in the limit of a small nonlinear inductance as
\begin{multline}
	\Delta f_n = \frac{n}{2\sqrt{C\left(L_{lin.}+L_{nl}\right)}}-\frac{n}{2\sqrt{L_{lin.}C}}  \\
    \approx -\frac{1}{2}\frac{n}{2\sqrt{L_{lin.} C}}\frac{L_k}{L_{lin.}}\left(\frac{I}		{I_*}\right)^2 = -\frac{1}{2}f_n\frac{L_k}{L_{lin.}}\left(\frac{I}		{I_*}\right)^2.
\end{multline}
If we now assume that current circulates due to energy stored in the pumped mode \(m\) (possibly with \(m=n\)) with an average number of photons \(\langle N \rangle_m = \frac{LI^2}{\hbar\omega_m}\) we obtain
\begin{equation} \label{eq:KerrCoeff}
	2\pi\Delta f_n \equiv \Delta \omega_n \approx  -\pi\frac{L_k}{L_{lin.}^2 I_*^2}\hbar \omega_n\omega_m \langle N \rangle_m \equiv K_{nm} \langle N \rangle_m.
\end{equation}
Where \(K_{nm}\) is termed self-Kerr (cross-Kerr) coefficient for the case \(n=m\) (\(n\neq m\))\cite{maleeva2018circuit}. Notice that \(K_{nm} = K_{11}\cdot nm\) for the case of a linear mode structure. Physically, this is a result of the linear dependence of pumping photon energy and shifted resonance frequency on mode number. We measured the self frequency shifts by varying the power of the probe field and measuring the resonance frequency shift for the different modes. The results of measurements from two different cool-downs are shown in Fig. \ref{fig:selfKerr}. At high pump powers the frequency shift is indeed negative, and seems to be linear in both stored energy and mode number (see inset in Fig. \ref{fig:selfKerr}) in agreement with Eqn. \ref{eq:KerrCoeff}.
Interestingly, at low energies we observe a \textit{positive} frequency shift when increasing the power, which is incompatible with the standard model of nonlinear kinetic inductance Eqn. \ref{eq:KI}.  The positive shift was more pronounced at the first cool-down but was also observed at the second cool-down. To our knowledge, the only known mechanisms for a positive frequency shift when using a single probe tone are broadband heating of the tunneling TLSs to temperatures of $T\gtrapprox 0.45 \frac{\hbar\omega}{k_B}$ \cite{gaoThesis}  or depopulation of thermal quasiparticles states by the microwave field \cite{deViseerEvidence,klapwijk2020discovery}. Since the last mechanism is relevant only at high temperatures \(T\gtrapprox 0.5 T_c\)  it seems to be irrelevant. Furthermore, since for our device \(\frac{\hbar\omega_1}{k_B} = 8\) mK, TLSs heating seems plausible. Another hint pointing to that direction is that it was recently reported \cite{mazin2020superconducting} that WSi has an anomalously high heat capacity at low temperatures, i.e. it may stay hot after it was heated by the radiation. We note that in addition to the weaker positive shift, at the second cool-down the quality factors were slightly higher than in the first one.  This might support the hypothesis of TLSs causing the shifts, since a new TLSs distribution  is generated upon thermal cycling the cryostat\cite{shalibo2010lifetime} which might also reduce the loss.  However, the positive shifts were not observed in cross-Kerr experiments (Fig. \ref{fig:crossKerr}, see below), implying a resonant effect and not a broadband heating. Another contradiction to a broadband heating model is the fact that the slope of the positive shift is \textit{larger} for the higher modes for which \(\frac{k_B T}{\hbar \omega_n}\) is \textit{smaller}. While frequency shifts resulting from a separate, detuned strong pump tone  were observed and explained as a consequence of asymmetric saturation of off-resonant TLSs \cite{KirshTLS, capelle2020probing} a single probe should result in symmetric saturation with no net frequency shift. The effect cannot be explained by a random asymmetry of TLSs frequencies due to their finite number \cite{KirshTLS, capelle2020probing} because the direction of the shift is positive for all probed modes. We note that positive shifts were observed by us also for devices with resonance frequencies of \(\sim 6-7\) GHz for which the temperature scale is much higher \cite{suppMaterial}. In addition, a positive frequency shift was also shown in the data of the high kinetic inductance NbN resonator reported in Ref. \cite{xing2020compact} for which the resonance frequency is \(\sim 6.8\)  GHz. In Ref. \cite{TiN} a positive shift in a \(\sim 1\) GHz TiN resonator was attributed to TLSs effects. Broadband heating of TLSs was used to explain positive frequency shifts of four harmonics of a \(\sim 2.14\) GHz Nb resonator in Ref. \cite{sage2011study}.  Further investigation of this effect is needed. Nevertheless, the presence of the positive frequency shift hinders the extraction of the bare self-Kerr coefficient.    
 
\subsection{Cross-Kerr nonlinearity and mode coupling}
\begin{figure} 
    \includegraphics[width=\linewidth,trim={0cm  0cm 13.125cm 0cm},clip]{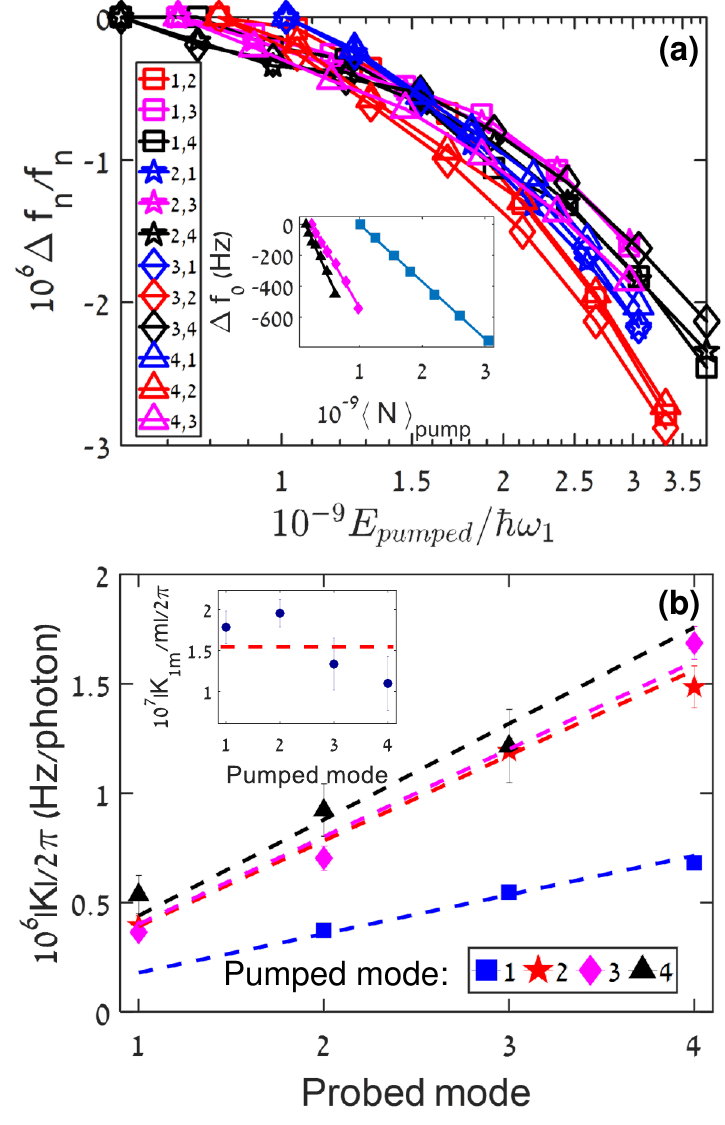}%
    \caption{Cross-Kerr frequency shifts.  (a) Differential frequency shift for different modes of the WSi device as a function of the stored energy in various pumped modes (in units of 1st mode photons).  In the legend (n,m) denotes probed mode = n, pumped mode = m. The inset shows linear fits to the frequency shifts of the 2nd mode. (b) Extracted cross-Kerr coefficients \(K_{nm}\)  when pumping mode \(m\) as a function of the probed mode \(n\). The dashed lines are a fits to \(K_{nm} \propto K_{1m} \cdot n \). The inset shows the fitted \(K_{1m}\) divided by pumped mode number \(m\)  as a function of  \(m\) which should equal \(K_{11}\). The dashed line is the average of \(K_{1m}/m\).  
    } \label{fig:crossKerr}
\end{figure}

In addition to the self-Kerr effects, in the second cool-down we also investigated the nonlinear shifts of a probed mode \(n\) when another mode \(m\) is strongly pumped. This was done by pumping at a constant frequency (the low-power resonance \(f_m\)) with various powers, while probing weakly around \(f_n\) in order to extract the frequency shift \(\Delta f_n\). Since the resonance frequency of the pumped mode will be shifted according to its own self-nonlinearity effects and its quality factor will also change, the stored energy \(E_{pumped}\) is not linear with the injected power. Therefore, we need to calculate \(E_{pumped}\) using the power-dependent quality factors and resonance frequencies extracted in the self-nonlinearity measurements.  The results are shown in Fig. \ref{fig:crossKerr}(a). As expected from Eqn. \ref{eq:KerrCoeff}, the differential shifts approximately coincide. Fitting the extracted cross-Kerr parameters for each pumped (probed) mode \(m\) (\(n\))  to \(K_{nm} = K_{1m}\cdot n\)  (see Fig. \ref{fig:crossKerr}(b)) we get \(K_{1m}\) for the first four modes. From the average of \(K_{1m}/m\) we estimate \(K_{11}/2\pi = -1.5(0.1) \times 10^{-7}\) Hz/photon which corresponds to \(I_* \approx 29\) mA, in agreement with transport measurements \cite{suppMaterial}. The variability of \(K_{1m}/m\) may be a result of the different overlaps of the spatial mode functions with local hot spots which causes an effective mode-dependent \(I_*\), or due to uncertainty in the calculation of \(E_{pumped}\).    
\section{Summary}
In conclusion, we have demonstrated the use of WSi as a high-kinetic-inductance material for both decreasing the footprint of a multimode resonator and for introducing both inter- and intra-mode nonlinearities to the device. The fabrication is straightforward and reproducible and yields a reduction of the footprint by a factor of \(\sim 2.9\) with standard lateral dimensions.  At high powers we observed a nonlinear behavior with a Kerr coefficient in the order of \(|K_{11}|/2\pi \approx 1.5\times10^{-7}\) Hz/photon for the first mode (172 MHz).  Further reduction of the footprint requires making \(\alpha\)  closer to \(1\). Since \(L_K \propto R\)  for \(T \ll T_c\) , where \(R\) is the normal resistance \cite{tinkham2004introduction}, this can be accomplished by a reduction of the cross section. Notice that \(L_g\) can be kept constant if the ratio between central strip width and the spacing to ground doesn't change \cite{Goppl2008}.  As can be seen from Eqn. \ref{eq:KerrCoeff} this will also increase the magnitude of the Kerr coefficient.  In addition, the Kerr coefficient can be increased by raising the fundamental frequency, since \(f_1 \sim \frac{1}{\sqrt{L_{tot}}}\) and therefore \(K_{11} \propto \left(f_1\right)^4\) at the limit of \(\alpha \approx 1\) of Eqn. \ref{eq:KerrCoeff}.  The footprint of the device can also be reduced independently from the nonlinearity by increasing its capacitance at the cost of a larger loss but with an impedance closer to 50 \(\Omega\) as demonstrated recently with hybrid Al/WSi devices \cite{goldstein2020four,goldstein2021compact}.  In this paper we used the nonlinear mode coupling for device characterization, but it may also be used for other purposes such as  measurement of one mode by monitoring another one \cite{buksIntermode, tancredi2013bifurcation} or even two-mode squeezing \cite{andersson2020squeezing} if the coupling is sufficiently strong when compared with losses.
\section{Acknowledgements}
We thank Dr. Tom Dvir for assistance with the transport measurements and the support of ISF Grant Nos. 963.19 and 2323.19.
\newpage
\providecommand{\noopsort}[1]{}\providecommand{\singleletter}[1]{#1}%

\end{document}


\title{Supplemental Material to "Linear and nonlinear properties of a compact  high-kinetic-inductance WSi multimode resonator"}

\author{Naftali Kirsh}
\affiliation{Racah Institute of Physics, the Hebrew University of Jerusalem, Jerusalem, 91904 Israel}
\author{Elisha Svetitsky}
\affiliation{Racah Institute of Physics, the Hebrew University of Jerusalem, Jerusalem, 91904 Israel}
\author{Samuel Goldstein}
\affiliation{Racah Institute of Physics, the Hebrew University of Jerusalem, Jerusalem, 91904 Israel}
\author{Guy Pardo}
\affiliation{Racah Institute of Physics, the Hebrew University of Jerusalem, Jerusalem, 91904 Israel}
\author{Ori Hachmo}
\affiliation{Racah Institute of Physics, the Hebrew University of Jerusalem, Jerusalem, 91904 Israel}
\author{Nadav Katz}
\affiliation{Racah Institute of Physics, the Hebrew University of Jerusalem, Jerusalem, 91904 Israel}

\date{\today}
\maketitle

\section{\(\lambda/4 \) WSi resonators}
\begin{figure}
    \includegraphics[width=\linewidth,trim={0cm 12.4cm 0cm 0cm},clip]{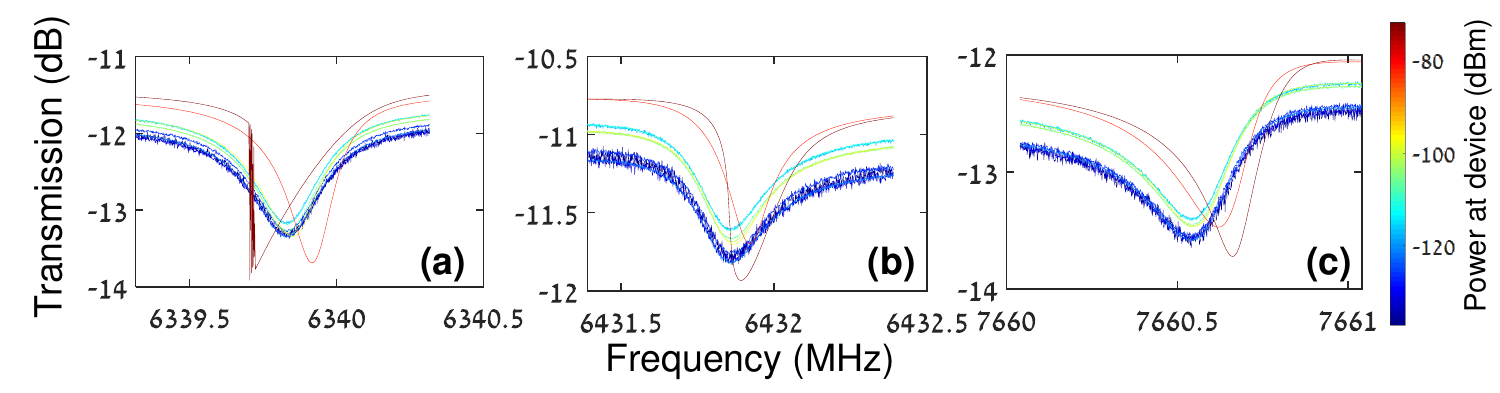} 
    \caption{Transmission measurements of different notch-type \(\lambda/4\) WSi resonators. In (a),(b) the third harmonic of two resonators is measured. In (c) the fifth harmonic of a third resonator is measured. The fundamental frequencies of the resonators in this device were outside the range of our amplifiers.
    } \label{fig:OriRes}
\end{figure}

As mentioned in the text, we also fabricated standard notch-type \(\lambda/4\) resonators from WSi films. The film thickness was 30 nm sputtered on sapphire. All resonators were coupled to a common feedline, through which the transmission was measured using a vector network analyzer. A few representative transmission spectra are shown in Fig. \ref{fig:OriRes}. The fundamental frequencies of the resonators in this device were outside the range of our amplifiers, hence we show here measurements of the third (Figs. \ref{fig:OriRes}(a) and \ref{fig:OriRes}(b)) harmonic of two resonators and the fifth (Fig. \ref{fig:OriRes}(c)) harmonic of another one. As can be seen, when increasing the power the resonance shifts clearly toward the positive direction. At the highest powers the resonance starts a negative shift (Fig. \ref{fig:OriRes}(b)) or even bifurcates (Fig. \ref{fig:OriRes}(a)). As discussed in the text, a positive frequency shift can be attributed to tunneling two-level systems heating, but notice that for these resonance frequencies the positive shift should start at a temperature of \(\sim 0.45 \frac{\hbar\omega}{K_B}\approx 137-166\) mK. Further investigation is needed.       

\section{Comparison to transport measurements}
\begin{figure}
	\includegraphics[width=\linewidth,trim={0cm  9.05cm 0cm 0cm},clip]{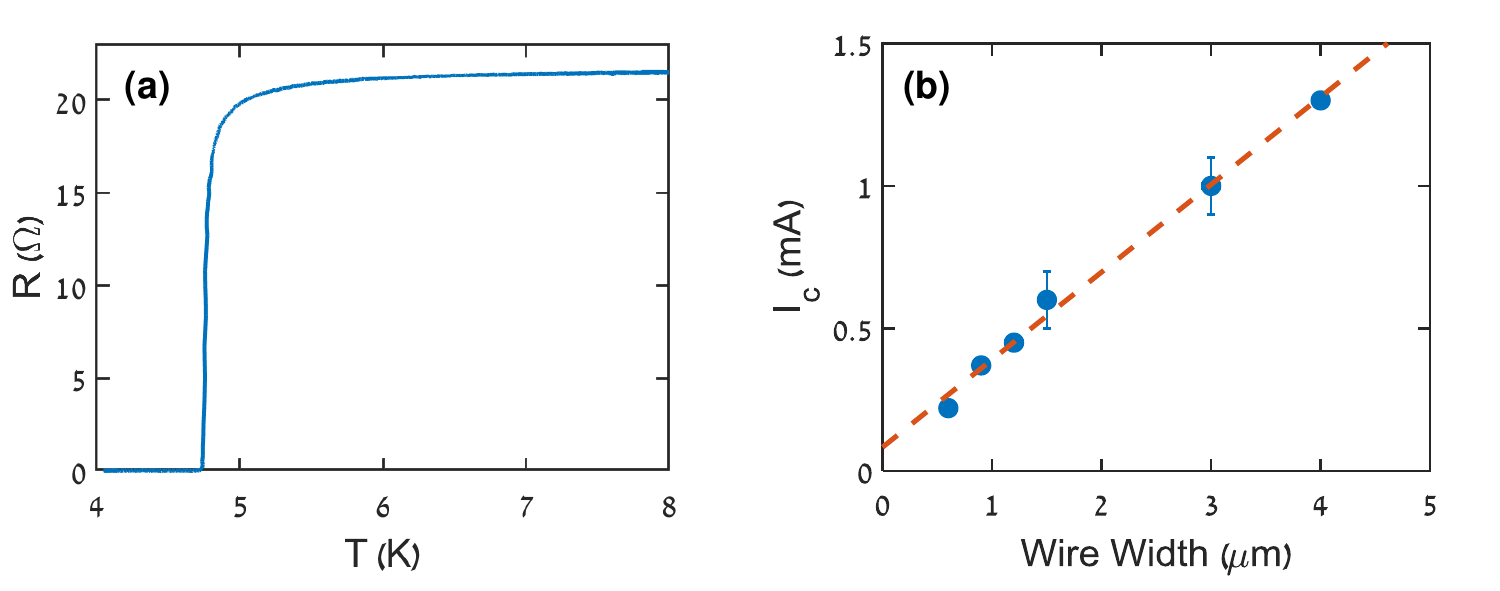} 
	\caption{Transport measurements of WSi devices. (a) Critical temperature measurement of a 30 nm thick 6 mm WSi chip. (b) Critical current measurement of a \(\sim\) 12 nm thick WSi chip with nanowires of various widths. The dashed line is a linear fit. }
	\label{fig:transport}
\end{figure}

Since the kinetic inductance depends on the normal state resistance \(R\) and energy gap \(\Delta\)  as \cite{tinkham2004introduction}
\begin{equation}
	L_K = \frac{\hbar R}{\pi \Delta}
\end{equation}when \(T \ll T_c\) , we can get an estimate for the expected kinetic inductance fraction \(\alpha\) from measurements of \(T_c\) and \(R\).  We performed a standard 4 wire measurement by bonding to a 6 mm chip of 30 nm thick sputtered film. Fig. \ref{fig:transport}(a) shows the critical temperature measurements from which we obtain \(T_c \approx 4.75\) K. From the normal resistance of the sample we can extract the resistance using the van der Pauw method from which we get \(\rho_n \approx  294 \mu\Omega\)-cm. Assuming a BCS superconductor with \(\Delta=1.764k_BT_c\) and using the dimensions of our device we get \(L_K \approx 414\) nH. The geometric inductance can be calculated using standard formulae\cite{Goppl2008} from which we approximate \(\alpha \approx 0.89\) in good agreement with our measurements.\newline

Another related transport parameter is the critical current \(I_c\) which is in the same order of \(I_*\) which sets the scale of the nonlinearity\cite{semenov2020effect}.  \(I_*\) can be extracted from the measured \(K_{11}\) and the calculated kinetic and geometric inductance using Eqn. (6) of the main text. Using the average value \(|K_{11}|/2\pi = 1.5\times10^{-7}\) Hz/photon we get \(I_* = 29\) mA.  We performed DC critical current measurements on a \(\sim 12\) nm thick WSi film with nanowires of various widths, see Fig. \ref{fig:transport}(b). Scaling to the dimensions of our device we can approximate \(I_* \approx 3I_c \approx 19.5\) mA with a reasonable agreement with the estimated \(I_*\) \cite{semenov2020effect}. 
\providecommand{\noopsort}[1]{}\providecommand{\singleletter}[1]{#1}%